\documentclass[fleqn,usenatbib]{mnras}
\usepackage{newtxtext,newtxmath}
\usepackage[T1]{fontenc}
\usepackage{ae,aecompl}
\usepackage{graphicx}	
\usepackage{amsmath}
\usepackage{amssymb}	
\title[Solar Coronal Heating by Gravity-Induced Resonant Emission]{Solar Coronal Heating by Gravity - Induced Resonant Emission}
\author[Antony Soosaleon]{
Antony Soosaleon,\thanks{E-mail: antonysoosaleon@yahoo.com (KTS)}
\\
Assistant Professor, School of Pure and Applied Physics, Mahatma Gandhi University, Kottayam, Kerala, India-686560\\
}
\begin{document}
\label{firstpage}
\pagerange{\pageref{firstpage}--\pageref{lastpage}}
\maketitle
\date{today}

\begin{abstract} 
	\textbf{Solar Coronal Heating is a Nonlinear Quantum Mechanical Phenomenon.}\\
		Corona is a powerful source of X-rays and ionisations \& emissions of such radiations are quantum mechanical process and energy levels corresponding to such emissions are highly unstable to order of femto-seconds. A linear method of energy transfer like collisional ionisation to such levels is practically not feasible. We have proposed \emph{\textbf{gire}} as the mechanism of this problem and discussed the physics, found it as a nonlinear process of multi photon absorption, similar to the multiple ionisation by ultra fast laser \citep{Antony2018}. Here, we are discussing the energy budget for ionisations of all the coronal elements and find \emph{\textbf{gire}} is sufficient.\\ 
	\textbf{Solar coronal heating balances the energy mass condition of the solar system, is a classical thermodynamical problem, but done by a nonlinear quantum mechanical process.}\\

\end{abstract}
\begin{keywords}
	Solar Coronal Heating, Collisional Ionisation, Uncertainty Principle, Multi Photo absorption, Atto Second transitions
	
\end{keywords}
\section{Introduction}

\subsection{Origin of the problem}
\label{S-Introduction} 
Solar coronal heating is a 75 years problem, started since 1939, when Eden and Bengt observed highly ionised iron. Later this was confirmed by several scientists which led to the observation of other highly ionised heavier elements. Average temperature of photosphere is around 5000 K and its pressure is of the order $10^{23}m^{-3}$, here the collision is dominant. Corona, the outer most layer of Sun and its pressure decreases to $10^{15}m^{-3}$, which is extremely vacuum and collision is negligible but, here we can observe all the electromagnetic (\emph{\textbf{em}}) radiations from highly the ionised states of heavier elements. This is very mysterious, since ionisation of heavier elements resulting to the emission of X-rays requires a temperature of million degree kelvin. By thermodynamical principle, heating to such a high temperature is not possible at this low pressure and this is called the Solar Coronal Heating Problem.\\

\begin{table}
	\caption{Constants \& Variables used for calculation of Ionisation energy}
	\label{T-simple}
	\begin{tabular}{clcc}     
		\hline 
		Symbol & Scientific name & Values used for calculation \\
		\hline \hline
		h & Planck's Constant & 6.626 $\times 10^{-34} Js^{-1}$ \\
		e & Electric charge & 1.602  $\times 10^{-19} C$ \\
		c & Velocity of light& 3 $\times 10^8 ms^{-1}$ \\
		m & Mass of electron & 9.1 $\times 10^{-31} kg$ \\
		M  & Mass of proton & 1.67 $\times 10^{-27} kg$ \\
		Z & Degree of ionisation & 1 (for initial coupling) \\
		W &  Mass number & referred from Periodic Table  \\
		$\gamma$ & Relativistic factor & calculated \& Table.2 \\
		$n_i$ & density of ions drift & $10^{9} kg m^{-4}$ \\
		$n_e$  & density of electrons  & $10^{9} kg m^{-3}$ \\ 
		$ \frac{n_i}{n_e}$ & Max. Field Deficit & $10^{7}m^{-1}$ for corona  \\    
		B & Magnetic field & 0.005 - 0.05 T \\
		g & Acc. due to gravity & 274 $ ms^{-2}$   \\
		$\theta$ & Angle between g \& B & depends on loop position \\
		\hline
	\end{tabular}
\end{table}

\subsection{Coronal Heating - Classical \& Quantum Mechanical}
"Solar Coronal heating" is a misleading term in its true sense of observational evidence. Emission of UV/X-rays from highly ionised heavier elements, does not requires a uniform heating of whole plasma, also it is insensible to consider a uniform collisional heating model when corona is highly vacuum and nonuniform in density \& field.  The observational evidences points out that the coronal emissions are preferentially localised and field oriented, in such case, uniform heating model is unreasonable but a particle resonance model will be suitable. \\

Even though coronal heating is a thermodynamical problem but basically it is a quantum mechanical process. Ionisation of heavier elements at high vacuum condition is fundamentally an atomic process, and such processes causes a high positive potential which resists further evaporation ions. In this way, the coronal heating balances the energy mass condition is a good reason to consider this problem as a thermodynamical one.\\
 \textbf{Coronal heating directly relates to the particle acceleration in solar wind which balances the energy mass conditions of solar system, which is a classical phenomenon but Ionisations and emissions are purely quantum mechanical processes. Thus coronal heating does a classical job in a quantum mechanical way.}

\subsection{Coronal Heating - Linear \& Nonlinear}
When an electron moves from higher to lower level, it emits energy in the form radiation and this is a very symmetric and linear quantum mechanical process, which balances the system and strictly obeys the quantum mechanical principles in conservation with time and energy. Thus absorption and emission are, a balanced act of energy flow, which is temporally a linear process, but in coronal heating, this is violated, the absorption is not symmetric with respect to the emission time. For any higher ionisation, the period of existing state cannot be less than the ionisation time. In other words, the energy required for further ionisation must be transferred to ion before getting de-excited from existing level. So the maximum time we can give for higher ionisation is the period of its existing level. If we analyse the periods of highly ionised states which results to the emission of  X-rays, they are in the range of femto seconds ($1 fs = 10^{-15} S$), for some even they are in the range of few atto seconds ($1 as = 10^{-18}S $). This is the main reason we put forward here that the ionisation of heavier elements in corona is not possible by collisional excitation or any other linear conventional method of energy transfer known in classical physics. This must be a nonlinear quantum mechanical process where the temporal compression is necessarily required to transfer such a huge amount of energy within a few femto/atto seconds.\\

\subsection{Coronal Heating  - Convergence of Fundamental Forces}
Sun is a perfect black body, source of all \emph{\textbf{em}} radiations and corona shows signatures of all these waves. We know the mechanism of gamma rays is associated with nuclear disintegration, X-rays and UV are by inner shell electronic transitions and radio waves are from gravity mode oscillations. Each of these process is unique, but all these radiations have been observed with MHD waves at same spatial point \citep{Antony2017b}. This indicates that these emissions are correlated and possibly from a single process at different time scale, and this is the reason.\\
 \textbf{\emph{gire} is the only solution, nonlinear, quantum mechanical and convergence all the fundamental variables suitable to analyse coronal emissions and we have already discussed and found that all observations converge in \emph{gire}} \citep{Antony2017b}.\\ 
\textbf{The most interesting part of \emph{gire} is that it connects the gravity with other fundamental forces}\\ (a detailed analysis is possible but that will change the motive of this paper, we reserve it for another discussion). \\
 
\section{Time, Energy \& Uncertainty Principle}
We have discussed the physics of \emph{\textbf{gire}} in the previous paper \citep{Antony2018}, the multiple ionisation is made possible by a bunch of ultra relativistic electrons at resonance condition and this was a nonlinear process similar to the pulsed laser ionisation. We have derived a formula for calculating the average energy absorbed by an ion per second. But, as we have discussed, the ionised states are highly unstable to the level of atto seconds and therefore, the energy must be calculated to each element for their corresponding ionisation time. This energy is compared with the ionisation energies of the elements taken from NIST atomic data base \citep{Kramida2018}, to find out the highly ionised states of the corresponding element, so that we can check whether \emph{\textbf{gire}} is effective for ionising the elements at coronal conditions. 

\subsection{Ionisation time}
Quantum Mechanics is a blessing to the humanity, as it can describe complicated phenomenons with simple mathematical equations. Here we make use of Heisenberg's intelligence for finding the time of ionisation. The rate absorption of photon energy in \emph{\textbf{gire}} is extremly high for which the ions undergo multiple ionisation and the whole process is controlled by the relativistic electrons, and therefore the uncertainty principle is the most suitable one for analysing the energy \& time of \emph{\textbf{gire}}.  

Let $\Delta t $ be the time of ionisation for $1^{st}$ to $2^{nd}$ ionised state of an ion, then $$(\Delta{E})(\Delta{t}) \geq \frac{\hbar}{2}$$. Where $\Delta E$ be the difference between $1^{st}$  \&  $2^{nd}$ ionisation energies for the observed coronal elements are taken from the NIST data centre. The ionisation time $\Delta{t}$  is calculated for all the coronal elements from He to Zn, are tabulated in the column 3 of table 3 (units are atto seconds).

\subsection{Role of Collisional Excitations}
Since the coupling of ions with electrons in \emph{\textbf{lhos}} is primary process for \emph{\textbf{gire}} and also the coupling starts at the lower part of corona where collision is dominated. Therefore the ionisation of neutral atoms by collision has an important role in the coronal heating. Solar spectrum has maximum Intensity at visible light, this is due to the transition at Balmer level (10.2 eV). This suggests the collisional excitations upto this level is continuous, and further only frequent, results to the emission of visible light. Due to this reason,  we keep the minimum value of energy as 10.2 eV for all elements with FIP $<10.2$eV, for the calculation of ionisation time $ \Delta t$.  

\begin{table}
	\caption{\textbf{Calculation of $T_i $, $T_e$ \& $\gamma $}}
	\label{T-simple}
	\begin{tabular}{cccccclc}     
		\hline 
		E &	W & FIP &  $T_i$ & $T_{ig}$ & $T_e$ & $\gamma$ \\
		\hline \hline
		H & 1.00		 &13.6 		 &5.02	   &5.02	     &8.272		    &1.03 \\
		He	& 4.00	    & 24.58	  & 5.28	 & 5.28	  	   & 9.07		   & 1.20	\\
		Li	& 6.94		 & 5.39		 & 4.62		& 4.62		  & 8.70		 & 1.15	\\
		Be	& 9.01		& 9.32		& 4.86		& 4.86	      & 9.01	   & 1.18	\\
		B	& 10.81		& 8.3		 & 4.81		  & 4.81		& 9.03		 & 1.21 \\
		C	& 12.01		& 11.26		& 4.94		& 4.94  	& 9.18		  & 1.26	\\
		N	& 14.01 	& 14.54		& 5.05		& 5.05	   & 9.33	     & 1.36	\\
		O	& 16.00		& 13.61		& 5.02		& 5.02	   & 9.35		& 1.38	\\
		F	& 19.00		& 17.42 	 & 5.13	   	  & 5.14	& 9.49		 & 1.52	\\
		Ne	& 20.18		& 21.56	 	& 5.22		& 5.22	 & 9.58		  & 1.64	\\
		Na	& 22.99		& 5.14  	& 4.60  	& 4.60	  & 9.14		& 1.40	\\
		Mg	& 24.31		& 7.64 		& 4.77 		& 4.77		& 9.30		& 1.42	\\
		Al	& 26.98		& 5.98		& 4.67		& 4.67  	& 9.25		& 1.46	\\
		Si	& 28.09		& 8.15		& 4.80		& 4.80	 	& 9.38		& 1.47	\\
		P	& 30.97		& 10.55		& 4.91		& 4.91		& 9.49		& 1.51	\\
		S	& 32.07		& 10.36		& 4.90		& 4.90		& 9.49	 	& 1.52	\\
		Cl	& 35.45 	& 13.01		& 5.00	   & 5.00		& 9.60		& 1.66	\\
		Ar	& 39.95		& 15.76		& 5.09	  & 5.09 	   & 9.70		& 1.82	\\
		K	& 39.10		& 4.34		& 4.53		& 4.53		& 9.26		 & 1.59	\\
		Ca	& 40.08		& 6.11		& 4.67		& 4.67		& 9.39		& 1.61	\\
		Sc	& 44.96 	& 6.56		& 4.71		& 4.70		& 9.45		& 1.66	\\
		Ti	& 47.88		& 6.83		& 	4.72	& 4.72		& 9.49		& 1.69	\\
		V	& 50.94		& 6.74		& 4.72		& 4.72		& 9.50		 & 1.73	\\
		Cr	& 52.00		& 6.76		& 4.72		& 4.72		& 9.51		& 1.73	\\
		Mn	& 54.98		& 7.43		& 4.76		& 4.76		& 9.56		& 1.76	\\
		Fe	& 55.84		& 7.9		 & 4.79		  & 4.79	  & 9.58	  & 1.76	\\
		Co	& 58.93		& 7.86		& 4.78		& 4.78		 & 9.60		& 1.79	\\
		Ni	& 58.69		& 7.63		& 4.77		& 4.77		 & 9.59	  	& 1.79	\\
		Cu	& 63.55		& 7.72		& 4.78		& 4.78		& 9.61		& 1.84	\\
		Zn	& 65.39		& 9.39		& 4.86		&  4.86 	& 9.68		& 1.86	\\
		\hline
	\end{tabular}\\
	\textbf{Information regarding columns:} \\
	1:	E- Elements Notation \\
	2:  W - Mass number  in a.m.u \\
	3:	FIP- First Ionisation Potential (eV)\\
	4:	$T_i$ ion temperature in log$_{10}$ scale, corresponding to FIP  ($\frac{3}{2}KT = eV$) \\
	5:	$T_{ig}$ ion temperature calculated by $T_i = \frac{Z^2}{W}\frac{\gamma m}{M}T_e $\\
	6:	$T_e$ electron temperature coupled with ions of $T_{ig}$ \\
	7:	$\gamma$  - relativistic factor corresponding $T_e$ \\
\end{table}
\begin{table}
	\caption{\textbf{Energy \& Ionised States for $B = 50G$ \& $(\frac{n_i}{n_e}) = 10^7$,}}
	\label{T-simple}
	\begin{tabular}{cccccclc}     
		\hline 
		E &1 IP & 2 IP & ${\Delta{E}}$ & ${\Delta{t}}$ & $E_{ob}$ &  $E_{Tx}$  & $X_x$  \\
		\hline \hline
		H 	&13.6 	&	--	 		& -- 		      &    --  	 &	--         & --            & --	\\
		He	& 24.58	  & 54.41     & 29.85 	  & 11.05  &  669.33 &  693.91 &  -- \\
		Li	 & 5.39		 &  75.64 	& 65.44     & 5.03  &   183.27 & 193.47 &-- \\
		Be	& 9.32		&  18.21  	& 8.01  	 & 41.16   &  1125.8& 1136  & -- \\
		B	& 8.3		&   25.15 	 & 14.95 	 & 22.05  &  490.14 & 500.34 & --\\
		C	& 11.26		& 	24.38 	& 13.12 	& 25.13  & 483.33 & 494.59   & --\\
		N	& 14.54		& 29.60  	& 15.06	   & 21.89  &  334.10 &  348.64  & $N_{6}$  \\
		O	& 13.61		&  35.12  	&  21.50 	& 15.34	  &  202.00 &  215.61 & $O_{7}$  \\
		F	& 17.42 	&  34.97 	 &  17.55    &18.80   &  189.27  & 206.69  &$F_{8}$  \\
		Ne	& 21.56	   &  40.96   & 19.40	  &  17.00	 & 149.35& 170.91  & $Ne_{7}$ \\
		Na	& 5.14  	& 47.28 	& 37.08   & 8.891   &  80.31  &  90.51   & $Na_{4}$  \\
		Mg	& 7.64 		&  15.03 	& 4.83    & 68.26   &  574.93  &  585.13  & $Mg_{11}$  \\
		Al	& 5.98		& 18.82 	& 8.62   	& 38.25   &  282.26  & 292.46   & $Al_{9}$  \\
		Si	& 8.15		&  16.35 	&	6.15	 & 53.61	& 377.50  & 387.7   & $Si_{10}$  \\
		P	& 10.49		&  19.77  	& 9.28		& 35.53   & 221.07  & 231.56   & $P_{7}$  \\
		S	& 10.36		&  23.34   & 12.98 	  & 25.40	&  151.62 & 161.99   &$S_{7}$  \\
		Cl	& 12.97		&  23.81	& 10.8 		& 30.42  &  150.83   & 163.84    & $Cl_{8}$   \\
		Ar	& 15.76		& 27.62	   & 11.86    & 27.80	&  111.17   & 126.93  & $Ar_{8} $   \\
		K	& 4.34		&  31.62 	& 21.42 	& 15.39	 & 71.975  & 82.175   &$K_{6}$    \\
		Ca	& 6.11		& 	11.87 	& 1.67        & 197.43  &  889.55 & 899.75  & $Ca_{16}$  \\
		Sc	& 6.56		& 	12.80  & 2.6    	& 123.95  &   483.31  &  493.51   & $Sc_{12}$  \\
		Ti	& 6.83		& 	13.58  	& 3.38  	& 97.54   &  350.47  &  360.67  & $Ti_{13}$   \\
		V	& 6.75		& 14.63    & 4.43      & 74.42   & 245.52 & 255.72 & $V_{12}$   \\
		Cr	& 6.77		& 16.48 	&  6.29 	& 52.42   & 169.68  & 179.88  &  $Cr_{8}$   \\
		Mn	& 7.43	   &  15.63    & 5.43 	  & 60.61  &  182.44 & 192.65	  & $Mn_{8}$   \\
		Fe	& 7.9		 &  16.20    & 6 .00  	& 54.95    &  162.56   & 172.76   & $Fe_{9}$  \\
		Co	& 7.88		& 17.08	 	& 6.88 		 & 47.92   & 132.08  & 142.28   & $Co_{8}$   \\
		Ni	& 7.64		& 18.17  	  & 7.97  	& 41.37	 & 114.49 & 124.69 & $Ni_{7}$  \\
		Cu	& 7.73		& 20.29 	& 10.09   & 32.68   & 81.26  & 91.46  & $Cu_{6}$  \\
		Zn	& 9.39		& 17.96 	& 7.76 		& 42.49  &  101.57	& 111.77 & $Zn_{7}$  \\		
		\hline
	\end{tabular}\\
	\textbf{Details of columns :} \\
	1:	E- Elements Notation \\
	2:  1 IP - First Ioisation potential (eV) \\
	3:	2 IP - Second Ionisation Potential \\
	4:	${\Delta{E}}$ =  2 IP - 1 IP \\
	5:	Ionisation time, ${\Delta{t}}$ =  $\frac{h}{4\pi}\frac{1}{\Delta{E}} $ \\
	6:	Energy absorbed (eV) by ion $E_{ob}$ = $ Es^{-1} \times {\Delta{t}}$ \\
	7:	Total Energy of the ion $E_{Tx}$  =  $E_{ob} + 1IP$ \\
	8: $X_x$ - Ionised state of element (no. of electrons removed is x-1) \\
	* \textbf{$X_x$ is found by comparing $E_{Tx}$ with Ionisation energies of corresponding element taken from NIST atomic data base \citep{Kramida2018}} \\
	
	\end{table}

\subsection{Ionisation Energy by \emph{gire}}

The energy absorbed by an ion per second is given by \citep{Antony2018}
\begin{equation}
Es^{-1} = (\frac{h c^2 e^2 }{4{\pi}^2 m M}) (\frac{Z}{\gamma W} ) (\frac{n_i}{n_e} \frac{B^2}{gSin\theta}) 
\end{equation}

The above formula has numerous constants \& variables, and we have collected them into 3 brackets. The first bracket contains constants and the second has variables depend on element for which the ionisation energy to be calculated. The third bracket is sensitively depend on the spatial and temporal conditions of corona where the emissions are to be analysed observed. 

\textbf{Variables depend on Element:} There are 3 variables in the $2^{nd}$ bracket; Z, degree of ionisation; W, mass number; $\gamma$, relativistic factor of coupling electrons with the corresponding ion. \\

\textbf{Degree of Ionisation Z:} This is the ionised state of atoms at the initial stage of coupling, normally all atoms are singly ionised state so that Z is taken as 1, and this is the possible ionisation of the atoms by collisional excitation at photosphere/transition region. But, there may be certain situation where double ionisation is possible for low FIP elements but here we are considering only singly ionised condition. \\

\textbf{Mass Number W:} Mass for each element is taken from the periodic table.\\

\textbf{Relativistic factor, $\gamma$:} depends on the energy of coupling electrons which is unique for each elements and it must be separately calculated. We know lower hybrid coupling is possible only when the Larmour radii of electrons and ions are same and imposing this condition, we have developed a formula connecting ion and electron temperatures ($T_i  \& T_e$ ) and their masses given as,

\begin{equation}
T_i = \frac{Z^2}{W}\frac{\gamma m}{M}T_e
\end{equation}  

Here, $T_i$ is calculated for each elements using the formula $\frac{3}{2}KT = eV$ with their corresponding first ionisation potentials assuming $Z = 1$. The coupling temperatures of electrons and $\gamma$ are calculated using the same procedure as in the paper \cite{Antony2017b}, the values are listed in the table 2. 

\subsection{ Field Deficit ($\frac{n_i}{n_e}$), Magnetic Field (B) \& $\theta$}
The $3^{rd} $ bracket contains the variables depend on the coronal conditions, for which they are highly fluctuating, so the values of these parameters should be taken from the observed coronal datas.\\

\textbf{Field Deficit ($\frac{n_i}{n_e}$)} : The mystery of coronal heating is hidden in this term and we have discussed the mechanism of this term thoroughly, how the electrons are lost at resonance condition also how it is amplified by the coronal loop oscillations. The decrease in electron density has been observed by several scientists \citep{Spadaro1990a, Doyle1998, Erdelyi1998, Chae1998, Esser1999, Doschek2001} which is of the order $10^7$. But this value can go upto $10^9$ in weak \textbf{B} region like corona if we calculate with respect to the elemental abundance.  Since all the electrons coupled in lower hybrid oscillations are accelerated out at resonance condition and also the transitions are possible at any level of field deficit.  Observations reveal coronal abundance for higher elements (except He) relative to hydrogen are lower with an order $10^6$. This means the density of heavier elements must be the order of $10^9 m^{-3}$ (average density of corona is $10^{15}$, which is less by a factor of $10^6$ will give $10^9$). Therefore, by charge neutrality condition when all the electrons are accelerated, then the field deficit can go upto this value,  but we keep the observed value for field deficit i.e., $10^7$.\\

\textbf{Magnetic field (B)}: The most deciding factor in the energy formula is B, since this is the only variable which contribute square times to the energy value. This is a very attractive reason since corona very magnetic and most of the coronal emissions are field based. The \emph{\textbf{lh}} coupling is possible only with ionised atom and we know the collisional ionisation is effective in the photosphere/ transition regions which is the lower part of corona. Here both particle and field pressure are effectively high, therefore we should consider the magnetic field corresponding to the photosphere/transition region. By Alfvenic wave prediction B ranges from 50 to 500G \citep{Aschwanden2005} and by gyro emission field has the values of 500 to 2500 G \citep{Doschek2001}. \emph{\textbf{gire}} is a gyro emission so, we can choose a high range, but choosing a higher magnetic field makes the job easier, also in consideration to quite sun, we choose low range 50 to 500 G.\\

\textbf{Angle between g \& B ($\theta$)}: The coronal B are in the form of loops and coupling starts at the lower part of corona where the angle between g and B must be very small and therefore we keep an average value for $\theta$ as $30$, but for corona and highly ionised form the value may go upto 90.

\section{Results \& Discussion} 

\subsection{Ionisation of Elements}The table 3 gives details of our calculation : the 6th column gives the approximate values of absorbed energy during ion cyclotron resonance and 7th column gives total energy of the ion (sum of absorbed energy with the collisional energy 10.2eV). The last column gives the ionised state of ions at this energy value.  Here we have taken the maximum value of field deficit, $(\frac{n_{\circ}^{'}}{n_{\circ}}) = 10^7$, for calculating the energy, it is to find out, whether \emph{\textbf{gire}} is effective in ionising all the coronal elements. But the transition is possible at any value of field deficit, therefore it is more descriptive if we go for calculating the energy absorption for various values of field deficit. 

Now second important variable in the corona is B,  we have calculated the energies and ionised states are tabulated for 100, 250 \& 500G shown in table 4. We find the lighter elements upto carbon has been disintegrated out and only heavier are sustaining at higher field regions. This is in resonance with observations that the corona reflects more abundance with mass to charge ratio (A detailed discussion is possible but reserved). Here also we have calculated the energy value for very high deficit to see the maximum possible ionisation. \\

In the above calculations we fixed the value of $\theta$ as 30 degree, for this, we said the reason that coupling is taking place at lower part of corona where the loop angle is very low. This is an assumption which need not be correct always since $\theta$ is unique for each emission.  Here we are doing an average  calculations for analysing to find out whether \emph{\textbf{gire}} is effective or not.  \\

The energy formula has several variables which are sensitively depend on the coronal conditions and we have done for an average values, but the most important variable is the result itself, i.e., the degree of ionisation Z, increases continuously when the ion start absorbing the energy, and this will further increases the rate absorption. While Z increases and also the value of $\gamma$ decreases marginally even though it is very small. So the variation in both Z and $\gamma$ has to be considered. But here we assume the energy transfer by multi photon absorption and multiple ionisation are one short process, otherwise an analysis is possible by fixing each ionisation energy.

\begin{table}
	\caption{\textbf{Comparison of ionisation for various B at $(\frac{n_i}{n_e}) = 10^7$ }}
	\label{T-simple}
	\begin{tabular}{cccccclc}     
		\hline 
		E & E(100) & $X_{x(100)}$ & E(250) & $X_{x(250)}$& E(500) & $X_{x(500)}$  \\
		\hline \hline
		O	& 0.83	&  $O_{8}$ &5.08 & -- & --& --    \\
		F	& 0.77	& $F_{8}$	& 4.75 & -- & --& --  \\
		Ne	& 0. 62 & $Ne_{9}$	& 3.76 & -- & --& --  \\
		Na	&  0.33   & $Na_{10}$& 2.02 & -- & --& --  \\
		Mg	& 2.31 & -- & --& --& --  & ---  \\
		Al	 & 1.14 & $Al_{11}$ & 7.07 & --& --  & --  \\
		Si	 &1.52 & $Si_{13}$ & 9.45 & --& --  & --  \\
		P	 & 0.90 & $P_{14}$  &5.53 & --& --  & --  \\
		S	&  0.62 &  $S_{13}$  & 3.80 & -- & --  & --  \\
		Cl	 &  0.62 & $Cl_{13}$ & 3.77  & $Cl_{17}$&--&--   \\
		Ar	&  0.46  & $Ar_{10} $  &  2.80 & $Ar_{17} $ & 11.12 & $Ar_{18}$   \\
		K	 &  0.30 &$K_{10}$   &  1.81 &$K_{18}$ & 8.75& $K_{18}$ \\
		Ca	 & 3.60 & $Ca_{19}$  & 22.26 &-- & -- & --  \\
		Sc	 &  1.94 &  $Sc_{20}$  & 12.09 &  -- &  --&--\\
		Ti	  & 1.41 &  $Ti_{20}$ &  8.78 & -- &  38.88 &--  \\
		V	  & 0.99 & $V_{16}$ &  6.15 & $V_{22}$  & 27.71& --    \\
		Cr	 & 0.69 &  $Cr_{15}$   & 4.25 &  $Cr_{23}$   & 19.03& $Cr_{23}$   \\
		Mn & 0.74 & $Mn_{16}$ &4.57& $Mn_{24}$ & 19.91& $Mn_{24}$   \\
		Fe	 &0.66&  $Fe_{17}$ &  4.07 & $Fe_{25}$  & 17.42 & $Fe_{25}$    \\
		Co	 & 0.54 & $Co_{17}$  & 3.31 &$Co_{26}$  & 14.14 & $Co_{24}$   \\
		Ni	 & 0.47 &  $Ni_{15}$ &2.87 &   $Ni_{27}$ & 12.41&$Ni_{23}$   \\
		Cu	 & 0.34 & $Cu_{12}$  & 2.03& $Cu_{18}$  & 8.85& $Cu_{22}$  \\
		Zn	&  0.42 & $Zn_{13}$  &  2.55 & $Zn_{17}$ & 10.43 &$Zn_{23}$ \\
		\hline
	\end{tabular}\\

	\textbf{Details of columns :} \\
1:	E- Elements Notation \\
2:  E(100) is sum of energy absorbed at B=100G and FIP (keV)\\
3: $X_{x(100)}$ is the ionised state of elements at E(100) \\
4: E(250) energy at B = 250G (keV)\\
5: $X_{x(250)}$ is the ionised state of elements at E(250) \\
6: E(500) energy at B = 500G (keV)\\
7: $X_{x(500)}$is the ionised state of elements at E(500)\\

\end{table}

\subsection{Conclusion}
 Solar coronal heating is a quantum mechanical problem but it seems to be a thermodynamical one which balances the energy mass flow. Emission of electromagnetic radiations are quantum mechnical process.  Highly ionised states of heavier elements results to the emission of X-rays are highly unstable to the order of atto seconds. Collisional excitations in this period of time is not feasible, a nonlinear mechanism of temporal compression is required in the energy transfer process to ionise the heavier elements to this higher degree. \emph{\textbf{gire}} does the job beautifully and the physics behind this process has been found as multi photon absorption which is a nonlinear quantum mechanical process. \\
\textbf{Theory of \emph{gire} is developed by classical fluid equations but it gives out the solution for frequency of resonant emission, which is quantum mechanical but the physics behind is a nonlinear process of multi photon absorption}\\

\textbf{Acknowledgment}

I am thankful to my wife Sunitha Ambrose, for her constant support in mat-lab programming and to my children, Filumena, Mary Cross \& John Flame for helping me in the calculations. \\
\textbf{I am grateful to NIST atomic data base \& Team,  without the values of Ionisation energies, this analysis is not possible.}\\

\textbf{This work is done under UGC Research Award Scheme (2016-2018)}.

\end{document}